\documentclass{article}

\usepackage{arxiv}

\usepackage[utf8]{inputenc} 
\usepackage[T1]{fontenc}    
\usepackage{hyperref}       
\usepackage{url}            
\usepackage{booktabs}       
\usepackage{amsfonts}       
\usepackage{nicefrac}       
\usepackage{microtype}      
\usepackage{lipsum}
\usepackage{graphicx}
\usepackage{hhline}
\usepackage{color}
\graphicspath{ {./images/} }
\usepackage[LGRgreek]{mathastext}

\title{Linguistic and Audio Embedding-Based \\Machine Learning for Alzheimer’s Dementia and \\Mild Cognitive Impairment Detection:\\ Insights from the PROCESS Challenge
}

\author{
 Adharsha Sam Edwin Sam Devahi  \\
  Singapore University of Technology and Design \\
   \And
 Sohail Singh Sangha  \\
  Singapore University of Technology and Design \\
   \AND
   Prachee Priyadarshinee \\
   Singapore University of Technology and Design \\
   \And
   Jithin Thilakan \\
   Hochschule für Musik Detmold \\
   \And
   Ivan Fu Xing Tan \\
      Singapore University of Technology and Design \\
 \\
   \And
   Christopher Johann Clarke  \\
   Singapore University of Technology and Design \\
      \And
   Sou Ka Lon \\
   Singapore University of Technology and Design \\
   \texttt{kalon\_sou@sutd.edu.sg} \\
   \And
   Balamurali B T  \\
   Singapore University of Technology and Design \\
   \texttt{balamuralibt@gmail.com} \\
    \And
   Yow Wei Quin\\
   Singapore University of Technology and Design \\
   \texttt{quin@sutd.edu.sg} \\
     \And
 Chen Jer-Ming   \\
  Singapore University of Technology and Design \\
  \texttt{jerming\_chen@sutd.edu.sg} \\
}

\begin{document}
\maketitle
\begin{center}
  {*All authors contributed equally to this work and are listed in no particular order.}  
\end{center}
\begin{abstract}
Early detection of Alzheimer’s Dementia (AD) and Mild Cognitive Impairment (MCI) is critical for timely intervention, yet current diagnostic approaches remain resource-intensive and invasive. Speech, encompassing both acoustic and linguistic dimensions, offers a promising non-invasive biomarker for cognitive decline. In this study, we present a machine learning framework for the PROCESS Challenge, leveraging both audio embeddings and linguistic features derived from spontaneous speech recordings. Audio representations were extracted using Whisper embeddings from the Cookie Theft description task, while linguistic features—spanning pronoun usage, syntactic complexity, filler words, and clause structure—were obtained from transcriptions across Semantic Fluency, Phonemic Fluency, and Cookie Theft picture description. Classification models aimed to distinguish between Healthy Controls (HC), MCI, and AD participants, while regression models predicted Mini-Mental State Examination (MMSE) scores. Results demonstrated that voted ensemble models trained on concatenated linguistic features achieved the best classification performance (F1 = 0.497), while Whisper embedding–based ensemble regressors yielded the lowest MMSE prediction error (RMSE = 2.843). Comparative evaluation within the PROCESS Challenge placed our models among the top submissions in regression task, and  mid-range for classification, highlighting the complementary strengths of linguistic and audio embeddings. These findings reinforce the potential of multimodal speech-based approaches for scalable, non-invasive cognitive assessment and underline the importance of integrating task-specific linguistic and acoustic markers in dementia detection.
\end{abstract}


\section{Introduction}

The early detection of dementia, particularly Alzheimer's Disease (AD) and Mild Cognitive Impairment (MCI), remains a critical challenge in clinical neuroscience \cite{brookmeyer2007forecasting,nichols2022estimation}. Speech, as a complex and readily accessible biomarker, holds significant promise for non-invasive assessment of cognitive decline \cite{meilan2014speech,priyadarshinee2023alzheimer,bt2024performance}. Leveraging advances in signal processing and predictive modeling, this study introduces an analytical framework designed to differentiate between Healthy Controls (HC), individuals with MCI, and those diagnosed with AD, using speech recordings. The present work builds upon the foundation laid by previous challenges such as ADReSS \cite{luz2021alzheimer}, ADReSSo \cite{luz2104detecting}, and ADReSS-M \cite{luz2024overview}, which have investigated automated detection of AD in various challenging scenarios, including limited training data and cross-lingual evaluation. However, unlike these previous initiatives, which focused primarily on the binary classification of AD versus HC, this study based on the PROCESS challenge expands the scope to include the detection of more subtle cognitive impairments associated with MCI. This expansion represents a notable departure that enables a more nuanced assessment of cognitive decline \cite{10889017}. This study employs a sophisticated signal processing pipeline and advanced machine learning algorithms to extract and analyze relevant acoustic and linguistic features from speech samples. By expanding the scope to include MCI, in addition to AD and HC, our goal is to develop a more nuanced and clinically relevant tool for the early identification and quantitative scoring of dementia-related pathologies.

\section{Process Challenge - Demographics, Data, Objective and Evaluation}

\subsection{Demographics}
\subsubsection{Training and Development Data}
A total of 157 participants were included in this study based on PROCESS challenge, forming the core dataset for our investigation into speech-based dementia detection. The age distribution of the participants displayed a wide range, ranging from 23 to 94 years, demonstrating a diverse pool of participants. The mean age was 65.7 years with a standard deviation of 12.3, with a median age of 66 years, indicating a relatively symmetrical distribution despite the wide range. Regarding gender, the participant pool consisted of 75 males,  81 females and 1 other, indicating a near-balanced gender representation. The diagnostic group included 59 participants with MCI, 82 Healthy Controls (HCs), and 16 participants diagnosed with AD. For the purpose of model development and evaluation, the data set was divided into 117 training samples (44 MCI, 61 HCs, 12 AD) and 40 development samples (15 MCIs, 21 HCs, 4 AD) \cite{10889017}. 

The availability of Mini-Mental State Examination (MMSE) scores for a subset of 69 participants (53 from training, 16 from development) offers crucial insights into the cognitive status of this cohort. With a mean score of 27.36 (SD = 2.47), and a median of 28, the data suggests that, on average, these participants exhibited cognitive function within the normal range. However, the observed MMSE range of 19 to 30 highlights the variability within the group, indicating that some individuals presented with mild cognitive impairment. This variability is essential for understanding the heterogeneity of cognitive profiles within the study population. The limited and sporadic availability of MMSE scores emphasizes the need for more comprehensive cognitive assessments in future studies to refine our understanding of the relationship between speech patterns and cognitive function.

\subsection{Data}
The data utilized in this study based on PROCESS challenge were derived from neuroscience research focused on dementia diagnosis and comprised audio recordings from three distinct elicitation tasks: Semantic Fluency (SF), Phonemic Fluency (PF), and Cookie Theft picture description (CTD) \cite{10889017,goodglass2001bdae}.
\begin{itemize}
    \item Semantic Fluency (SF): To access and retrieve semantic knowledge, participants were instructed to "Please name as many animals as you can in a minute." This task, similar to naming tasks commonly employed in cognitive assessments, primarily evaluates language abilities and naming skills, serving to detect potential issues in language comprehension and expression \cite{olmos2023phonological}.
    \item Phonemic Fluency (PF): To access phonological and lexical retrieval mechanisms, participants were asked to "Please say as many words beginning with the letter ‘P’ as you can in a minute. Any word beginning with ‘P’, except for names of people such as Peter, or countries such as Portugal." This task is designed to assess verbal fluency and executive functions related to language \cite{olmos2023phonological}.
    \item Cookie Theft picture description (CTD): Participants were presented with the "Cookie Theft" picture and asked to describe it. This task is intended to reflect various cognitive functions of the speakers, including language comprehension and memory. Further, this task helps in accessing linguistic functions such as semantic categories, referential cohesion, and language and speech structure \cite{cummings2019describing}.
\end{itemize}
These three tasks, each targeting different aspects of cognitive function, provide a comprehensive dataset for the development and evaluation of automated dementia detection algorithms.

\subsection{Objective}
The primary objective of this study was twofold: first,  a classification task aimed to accurately identify individuals with varying cognitive impairments and dementia, specifically distinguishing between HC, those with MCI, and those with AD, based solely on their audio samples; second, a regression task focused on predicting the MMSE score for each participant, providing a quantitative measure of cognitive function. These objectives address critical aspects of dementia detection and assessment, leveraging speech as a non-invasive biomarker to facilitate early and accurate diagnosis.

\subsection{Evaluation Metrics}
Performance for this study was evaluated using distinct metrics for the classification and regression tasks. For the classification task, which focused on distinguishing between HC,  MCI, and AD, macro-averaged metrics was utilized: Macro F1 Score, the harmonic mean of Macro Precision and Macro Recall, providing a balanced measure of overall performance, where Macro Precision, representing the average ratio of correctly predicted positive observations to total predicted positives and Macro Recall, representing the average ratio of correctly predicted positive observations to all actual positive observations.  For the regression task, which involved predicting MMSE scores, the Root Mean Square Error (RMSE) between the actual and predicted MMSE scores was used as the primary evaluation metric \cite{10889017,sokolova2006beyond}.

\section{Methodology}
This study employs a multimodal approach, generating text transcriptions from audio recordings to leverage the inherent multimodality of speech. The methodology is grounded in the demonstrated efficacy of multimodal analyses in AD prediction, where the integrated analysis of both acoustic and linguistic features has shown superior performance in dementia detection \cite{priyadarshinee2023alzheimer}.
\subsection{Data Processing}
\subsubsection{Audio Data Preprocessing}
Prior to feature extraction, a crucial preprocessing step was implemented to isolate relevant speech segments and eliminate non-speech portions from the audio recordings. This step was essential for focusing the analysis on the articulation characteristics of the participants, thereby minimizing the influence of extraneous noise and silence. We employed Silero-VAD, an open-source Voice Activity Detection model, to achieve this. Silero-VAD \cite{SileroVAD} was selected due to its demonstrated effectiveness and its training on extensive corpora spanning over 6000 languages. This extensive training enables the model to accurately identify speech segments across diverse linguistic and acoustic environments, ensuring that the subsequent feature extraction is performed on relevant speech data. While the aggressive removal of background silence and noise enhances the identification of relevant speech signals for modeling, it also results in the potential loss of pertinent acoustic information. For instance, subtle variations in pauses between utterances could reveal insights into the participant's cognitive state. Recognizing this limitation, we acknowledge that the preprocessed audio represents a trade-off between signal clarity and the preservation of potentially relevant acoustic cues. Therefore, future investigations might explore alternative VAD techniques or post-processing methods, such as extracting statistics from the removed silence, to mitigate this loss while maintaining the benefits of speech isolation.

Speaker diarization, the automatic segmentation of audio by speaker, was considered but not implemented. Initial analysis, validated with Whisper v3 and Pyannote \cite{bredin2020pyannote}, showed minimal interviewer speech. Given the predominantly single-speaker nature of the audio data, the potential benefits of diarization were deemed insufficient to justify the computational cost and potential for error introduction.

\subsubsection{Audio Features --- Whisper embeddings} 

This study employed Whisper embeddings for audio feature representation \cite{radford2023robust}. Whisper, a transformer-based encoder-decoder model, was pre-trained on a large-scale, multilingual dataset, enabling robust feature extraction across diverse acoustic conditions. The embeddings were derived from the model's encoder layers, specifically the penultimate layer's output, resulting in a 1280-dimensional, continuous vector representation of the input audio. This representation encapsulates both acoustic and phonetic information, as learned by the model during its extensive pre-training. The utilization of Whisper embeddings was motivated by their demonstrated efficacy in capturing complex acoustic patterns, facilitating downstream tasks such as speech recognition and speaker verification. Specifically, the study aimed to leverage the learned representations to discern subtle acoustic variations within speech that may correlate with cognitive decline, thus providing a refined input for subsequent classification and regression models.

Among the three elicitation tasks – CTD, SFT, and PFT – the CTD consistently demonstrated the most promising results from our preliminary analysis in terms of predictive performance for both the classification and regression tasks. This was evident during the training and validation phases, where models trained on CTD data exhibited superior accuracy in distinguishing between cognitive states and predicting MMSE scores. Consequently, subsequent analyses focused exclusively on the CTD task, utilizing Whisper embeddings derived from these recordings. A consistent degradation of model performance on the development set was observed when data from SFT and PFT were incorporated, regardless of whether the features were concatenated or fused sequentially with the CTD data. This empirical finding suggested that the inclusion of SFT and PFT data with the model chosen in this investigation, rather than enhancing the model's ability to discern relevant patterns, their inclusion introduced noise or obscured the salient features present in the CTD data. As a result, only the Whisper embeddings derived from the CTD task were utilized to evaluate the performance of the models on the final test dataset.

\subsection{Text Data Processing}
\subsection{Speech-to-Text Conversion}

 Several state-of-the-art automatic speech recognition (ASR) systems, including wav2vec \cite{baevski2020wav2vec}, DeepSpeech \cite{hannun2014deep}, Whisper v3, and CrisperWhisper \cite{wagner2024crisperwhisper}, were evaluated. While these deep learning models exhibited comparable transcription accuracy, CrisperWhisper was ultimately selected for its unique ability to accurately transcribe disfluencies, such as filler words (`um', `uh'). The inclusion of these linguistic markers is crucial, as they provide valuable indices of cognitive processes, including hesitation and lexical retrieval challenges, which are frequently observed in individuals with cognitive impairments.

\subsubsection{Text Features --- Traditional Linguistic Features}

For linguistic feature extraction, the features were selected to capture a range of syntactic and lexical characteristics that are associated with cognitive decline, based on previous research \cite{chapin2022finer}.  The feature set comprised the following: duration, pronoun ratio, percentage of definite pronouns and percentage of indefinite pronouns, total noun phrase rate, filler word rate, word count rate, active interaction, adverbial adjunct ratio, total clause rate, and adjunct clause ratio. These features aimed to provide a comprehensive linguistic profile of each participant, enabling the investigation of potential correlations between specific language patterns and cognitive status. The details of this feature set are listed in Table 1. These features were extracted using natural language processing techniques such as part-of-speech (POS) tagging and dependency parsing implemented with the SpaCy NLP library \cite{spacy2015industrial}.


\begin{table}[h]
\renewcommand{\arraystretch}{2.5}
 \centering
\caption{Linguistic Features}
\begin{tabular}{|l|c|}
\hline
\textbf{Feature}& \textbf{Formula}\\
\hline
$pronoun\_ratio$ & $\frac{pronoun\_count}{pronoun\_count +definite\_np\_count + indefinite\_np\_count}$\\

\hline

$percent\_definite$ & $\frac{definite\_np\_count}{pronoun\_count + definite\_np\_count + indefinite
\_np\_count}$\\

\hline
$percent\_indefinite$ & $\frac{indefinite\_np\_count}{pronoun\_count + definite\_np\_count + indefinite
\_np\_count}$\\

\hline
$total\_np\_rate$ & $\frac{pronoun\_count + definite\_np\_count + indefinite\_np\_count}{duration}$\\

\hline
$filler\_word\_rate$ & $\frac{filler\_word\_count}  {duration}$\\

\hline
$total\_word\_count\_rate$ & $\frac{total\_word\_count}  {duration}$\\

\hline
$active\_interaction$ & $\frac{actual\_word\_count}  {total\_word\_count}$\\

\hline
$adverbial\_adjunct\_ratio\_punct$ & $\frac{adverbial\_adjunct\_count}  {total\_sentence\_count\_punct}$\\

\hline
$adverbial\_adjunct\_ratio\_sentstruct$ & $\frac{adverbial\_adjunct\_count}  {total\_sentence\_count\_sentstruct}$\\

\hline
$total\_clause\_rate\_minimal^*$ & $\frac{total\_clause\_count\_minimal}  {duration}$\\

\hline
$total\_clause\_rate\_comprehensive$ & $\frac{total\_clause\_count\_comprehensive}  {duration}$\\

\hline
$adjunct\_clause\_ratio\_minimal$ & $\frac{adjunct\_clause\_count}  {total\_clause\_count\_minimal }$\\

\hline
$adjunct\_clause\_ratio\_comprehensive$ & $\frac{adjunct\_clause\_count}  {total\_clause\_count\_comprehensive }$\\

\hline
\hline
\end{tabular}
\end{table}

To comprehensively represent the linguistic characteristics of participant speech, a unified feature set was constructed. This set comprised 14 traditional linguistic features extracted from the transcribed text of each audio recording. Recognizing the potential for task-specific linguistic variations, data from all three elicitation tasks (CTD, SFT and PFT) were integrated. Specifically, the 14 features extracted from each task were horizontally concatenated, resulting in a single 42-dimensional feature vector per participant. This approach aimed to capture a holistic linguistic profile, reflecting potential variations in language production across three different cognitive tasks and providing a robust input for subsequent modeling.

\subsubsection{File-level vs Frame-level Features}

Overall, this study adopted a file-level feature analysis, ensuring a consistent number of features across all audio and text files enabling
the exploration of robust, albeit static, relationships between acoustic and linguistic patterns and cognitive status. This approach facilitated the application of both traditional machine learning algorithms and deep neural networks (DNNs). Specifically, the feature vectors derived from Whisper embeddings were aggregated at the file level and traditional linguistic analyses were aggregated at the participant level concatenating the features from three elicitations, representing each participant's speech and text data as a single, fixed-length vector respectively for subsequent audio and text modelling.

While frame-level features, which capture the dynamic temporal evolution of speech, offer potential advantages in modeling  time-varying acoustic changes \cite{priyadarshinee2023alzheimer}, their inclusion was deliberately avoided in this study. The decision was primarily driven by the hardware impracticality of processing sheer volume of time-series data coupled with the necessity for a rapid development and evaluation cycle within the PROCESS competition's timeframe.
\subsection{Machine Learning Models and Evaluation}

To investigate the relationship between extracted features (from both speech and the text) and cognitive status, this study explored a diverse set of machine learning models. The models included ensemble methods, specifically Random Forests, AdaBoost, and Gradient Boosting, which are known for their robustness and ability to handle complex data patterns \cite{kunapuli2023ensemble}. Additionally, Support Vector Machines (SVMs) were utilized, given their effectiveness in high-dimensional feature spaces. Deep Neural Networks (DNNs) were also incorporated to assess the potential of deep learning architectures in capturing intricate relationships within the transformation of the data.

For each model, hyperparameter optimization was conducted to identify the optimal parameter configurations. This process involved a grid search within a predefined parameter space, aiming to maximize model performance. Model evaluation was performed using a 5-fold cross-validation strategy on the training dataset. This technique provided a robust estimate of model performance by partitioning the training data into five subsets, iteratively training on four subsets and validating on the remaining one. The hyperparameters that yielded the best average performance across the cross-validation folds were then used to train the final model, which was subsequently evaluated on the independent development set.

In addition, ensemble voting techniques were investigated to potentially enhance predictive performance \cite{priyadarshinee2023alzheimer,bharati2022dementia}. Hard and soft voting strategies were explored. Hard voting involved taking the majority vote of the predicted class labels from multiple models, while soft voting combined the probability estimates from each model before making a final prediction. These voting techniques were applied to combine the predictions of multiple models trained on a given feature set, aiming to leverage the complementary strengths of different algorithms and improve overall accuracy and robustness.


For the classification task, Whisper embeddings from the CTD task yielded optimal development set results for audio, achieved by a soft-voted ensemble of Random Forest, AdaBoost, and DNNs trained on the training set. For text, a Random Forest classifier performed best using concatenated linguistic features from all three tasks (CTD, PFT, SFT).

For the regression task, a similar modeling approach to that of the classification task was employed across both text and audio features. A voting regressor, combining Random Forest, AdaBoost, and Gradient Boosting regressors, was utilized for both feature sets. This ensemble method was applied to the concatenated linguistic features for text and to the Whisper embeddings derived from the CTD task for audio.

\section{Results}

For this study based on PROCESS challenge, participants were permitted to submit up to three distinct model results for each task. This allowed for the evaluation of diverse modeling approaches, with a maximum of three submissions for the classification task and three submissions for the regression task, totaling six submissions per participant.
\subsection{Classifying cognitive impairment and dementia participants from healthy volunteers}
For the classification task, three distinct models were evaluated. Model 1, utilizing concatenated linguistic features from all three elicitation tasks (CTD, PFT, SFT) and a Random Forest classifier,  trained on the combined training and development set, achieved an F1 score of 0.497 on the test set. Models 2 and 3 employed Whisper embeddings from the CTD task exclusively, with a soft-voted ensemble of Random Forest, AdaBoost, and DNN classifiers. Model 2, trained solely on the training set, yielded an F1 score of 0.372, while Model 3, trained on the combined training and development sets, achieved an F1 score of 0.400. These results indicate that the linguistic feature-based model outperformed the Whisper embedding-based models in this classification task. Furthermore, the inclusion of the development set in training improved the performance of the Whisper embedding-based ensembles.

\subsection{MMSE Prediction}
For the MMSE prediction regression task, again three models were evaluated. Model 1, utilizing concatenated linguistic features from all three elicitation tasks (CTD, PFT, SFT) and a voting regressor comprising Random Forest, AdaBoost, and Gradient Boosting, trained on the combined training and development sets, achieved a Root Mean Squared Error (RMSE) of 2.915 on the test set. Models 2 and 3 employed Whisper embeddings from the CTD task exclusively, with the same voting regressor. Model 2, trained solely on the training set, yielded an RMSE of 2.957, while Model 3, trained on the combined training and development sets, achieved an RMSE of 2.843. These results indicate that the Whisper embedding-based model trained on both training and development data performed best in this regression task. Furthermore, the inclusion of the development set in training improved the performance of the Whisper embedding-based regressors.

\subsection{Comparative Performance Analysis among submissions in PROCESS Challenge}

In the PROCESS Challenge, our submissions were evaluated against 106 classification and 80 regression models. For classification, Model 1 (linguistic features, Random Forest) ranked $34^{th}$ (F1 0.497), while Models 2 and 3 (Whisper embeddings, soft-voted ensemble) ranked $90^{th}$ (F1 0.371) and $85^{th}$ (F1 0.400) respectively. Model 1 and 3 was trained on combined training/development data. The top classification score was 0.696. 

For regression, Model 3 (Whisper embeddings, voting regressor, combined training/development) ranked $15^{th}$ (RMSE 2.843). Model 1 (linguistic features, voting regressor) ranked $24^{th}$ (RMSE 2.915), and Model 2 (Whisper embeddings, voting regressor, training data only) ranked $29^{th}$ (RMSE 2.957). The top regression score was 2.459. 

\section{Conclusions}

This study demonstrates the viability of leveraging multimodal speech analysis—integrating linguistic features and audio embeddings—for the automated detection of Alzheimer’s Dementia and Mild Cognitive Impairment. Linguistic features derived from transcribed speech showed superior performance for multi-class classification of cognitive status, while Whisper embeddings provided stronger predictive power for MMSE regression. These complementary findings highlight that speech-based biomarkers capture distinct but convergent aspects of cognitive decline. Although our results ranked mid-range for classification and top-tier for regression within the PROCESS Challenge, they underscore the translational potential of speech-based machine learning systems as scalable tools for early screening in clinical and community settings. Future work should explore dynamic, frame-level acoustic features, integrate richer cognitive task designs, and validate models across larger and more diverse cohorts to enhance robustness and generalizability. Ultimately, combining linguistic and acoustic representations provides a promising path toward non-invasive, low-cost, and clinically meaningful assessment of dementia-related pathologies.

\section* {Acknowledgement}

This research is supported by the Ministry of Education Academic Research Fund Tier 1, Singapore, under its SUTD Kickstarter Initiative awarded to Prof W. Quin Yow (SKI 2021\_03\_11).

\bibliographystyle{unsrt}  
\bibliography{template}  


\end{document}